\documentclass[twocolumn]{aastex63}
\pdfoutput=1 

\usepackage{amsmath,amstext}
\usepackage[T1]{fontenc}
\usepackage{apjfonts, xspace, xcolor} 
\usepackage[figure,figure*]{hypcap}
\usepackage{xcolor}
\usepackage{csvsimple}
\usepackage{longtable}
\usepackage{xcolor}
\usepackage[colorinlistoftodos]{todonotes}
\usepackage{makecell}
\usepackage{float}
\usepackage{cleveref}
\usepackage{graphicx}
\usepackage{subfigure}
\usepackage{tabularx}
\usepackage{lipsum}
\usepackage{soul}

\def\NumDots{52}\def\DotLifetime{72}\def\DotLifetimeStd{81}\def\EUIintensity{32}\def\EUIintensityStd{22}\def\IRISintensity{134}\def\IRISintensityStd{113}\def\EUIsize{679}\def\EUIsizeStd{367}\def\IRISsize{788}\def\IRISsizeStd{309}\def\DotSpeed{37}\def\DotSpeedStd{34}

\shorttitle{Bright Dots and Coronal Plume Formation in Sunspot Penumbra}
\shortauthors{Weitz et al.}

\begin{document}

\title{Bright Dots and Coronal Plume Formation in Sunspot Penumbra}


\author[0000-0002-4401-2295]{Ayla Weitz}
\altaffiliation{CU Boulder Hale Graduate Fellow}
\affiliation{University of Colorado Boulder, 2000 Colorado Ave, Boulder, CO 80309, USA}
\affiliation{National Solar Observatory, 3665 Discovery Dr, Boulder, CO 80303, USA}
\affiliation{Lockheed Martin Solar and Astrophysics Laboratory, 3251 Hanover Street, Bldg. 203, Palo Alto, CA 94306, USA}
\affiliation{Bay Area Environmental Research Institute, NASA Research Park, Moffett Field, CA 94035, USA}

\author[0000-0001-7817-2978]{Sanjiv K. Tiwari}
\affiliation{Lockheed Martin Solar and Astrophysics Laboratory, 3251 Hanover Street, Bldg. 203, Palo Alto, CA 94306, USA}
\affiliation{Bay Area Environmental Research Institute, NASA Research Park, Moffett Field, CA 94035, USA}

\author[0000-0002-6116-7301]{Gianna Cauzzi}
\affiliation{National Solar Observatory, 3665 Discovery Dr, Boulder, CO 80303, USA}
\affiliation{Istituto Nazionale di Astrofisica, Osservatorio Astrofisico di Arcetri, Largo E. Fermi 5, I-50125 Firenze, Italy
}
\author[0000-0001-8016-0001]{Kevin P. Reardon}
\affiliation{University of Colorado Boulder, 2000 Colorado Ave, Boulder, CO 80309, USA}
\affiliation{National Solar Observatory, 3665 Discovery Dr, Boulder, CO 80303, USA}

\author[0000-0002-8370-952X]{Bart De Pontieu}
\affiliation{Lockheed Martin Solar and Astrophysics Laboratory, 3251 Hanover Street, Bldg. 203, Palo Alto, CA 94306, USA}
\affiliation{Rosseland Centre for Solar Physics, University of Oslo, P.O. Box 1029 Blindern, NO-0315 Oslo, Norway}
\affiliation{Institute of Theoretical Astrophysics, University of Oslo, P.O. Box 1029 Blindern, NO-0315 Oslo, Norway}

\correspondingauthor{Ayla Weitz}
\email{Ayla.Weitz@colorado.edu}

\begin{abstract}
Coronal plumes are narrow, collimated structures that are primarily viewed above the solar poles and in coronal holes in the extreme ultraviolet, but also in sunspots. Open questions remain about plume formation, including the role of small-scale transients and whether plumes embedded in different magnetic field configurations have similar formation mechanisms. 
We report on coordinated Solar Orbiter/Extreme Ultraviolet Imager (EUI), Interface Region Imaging Spectrograph, and Solar Dynamics Observatory observations of the formation of a plume in sunspot penumbra in 2022 March. During this observation, Solar Orbiter was positioned near the Earth–Sun line and EUI observed at a 5 s cadence with a spatial scale of 185 km pixel$^{-1}$ in the solar corona. We observe fine-scale dots at various locations in the sunspot, but the brightest and highest density of dots is at the plume base. Space-time maps along the plume axis show parabolic and V-shaped patterns, and we conclude that some of these dots are possible signatures of magneto-acoustic shocks. Compared to other radial cuts around the sunspot, along the plume shows the longest periods ($\sim 7$ minutes) and the most distinct tracks. Bright dots at the plume base are mostly circular and do not show elongations from a fixed origin, in contrast to jetlets and previously reported penumbral dots. We do not find high-speed, repeated downflows along the plume, and the plume appears to brighten coherently along its length. Our analysis suggests that jetlets and downflows are not a necessary component of this plume's formation, and that mechanisms for plume formation could be dependent on magnetic topology and the chromospheric wave field.
\end{abstract}

\section{Introduction}

Coronal plumes are bright, narrow, collimated structures that are primarily rooted inside coronal holes. They were first observed in the K-corona above the solar poles, but are also visible in the extreme ultraviolet (EUV), which allows them to be seen on the solar disk as well.
Coronal plumes are associated with open magnetic field lines extending into the heliosphere, meaning these features only have one visible footpoint. Understanding the formation and evolution of plumes provides insight into the topology of coronal magnetic fields and the potential mass flux supply for the solar wind \citep{Poletto_2015}. 

Several episodic phenomena have been proposed as the primary sources for forming and sustaining coronal plumes. These include jetlets \citep{Raouafi_2009,McIntosh_2010,Pant_2015, Uritsky_2021, Kumar_2022}, magnetic reconnection \citep{Wang_2016,Panesar_2018}, plume transient bright points \citep{Raouafi_2014}, and spicules \citep{Wilhelm_2000, Cho_2023}. 
Some of these authors have noted that precursors tend to recur with periods of three to 5 minutes, similar to the periods of the \emph{p}-mode oscillations in the photosphere.
Beyond these transient phenomena, plumes are known to host magnetohydrodynamic (MHD) waves \citep{DeForest_1998, Thurgood_2014}, and MHD modeling has suggested that upwardly propagating slow magnetosonic waves may contribute significantly to lower coronal heating via compressive dissipation \citep{Ofman_1999}. Additionally,
the dissipation of kinetic Alfv\'en waves has been suggested as a potential mechanism for localized plume heating \citep{Wu_2003}.

Although less common, coronal plumes are also observed in sunspots \citep{Foukal_1976,Maltby_1998,Doyle_2003}. Termed ``sunspot plumes,'' they are generally defined as extended, EUV-bright loops emanating from sunspot umbrae (and sometimes penumbrae) that do not clearly show a corresponding, opposite-polarity footpoint. A similar type of structure is the fan loop \citep{Schrijver1999, Ugarte-Urra_2011}, which shows a significant lateral widening above a bright and compact footpoint in the umbra or penumbra. Both features are associated with relatively cool loops (temperatures at or below 1 MK). 
Due to the vastly different magnetic field configurations between coronal holes and sunspots, it is an open question whether plumes in these regions have similar formation mechanisms. 
Previous studies found that inflows are a requirement for sunspot plumes to occur \citep{Brynildsen2001}, and these flows usually correspond with chromospheric and transition region bright dots at plume footpoints in the umbra \citep{Chen_2022}.

Whether or not bright dots are a necessary component of sunspot plumes, the appearance of bright dots in sunspots is common. Previous studies of penumbral dots find elongation and some movement along penumbral filaments in the transition region \citep{Tian_2014} and in the corona \citep{Alpert_2016}. Small-scale brightenings at coronal loop footpoints in sunspot umbrae and penumbrae have also been related to coronal rain \citep{Kleint_2014}. The penumbrae of sunspots also seem to be the site of numerous jetlike phenomena that can be detected in a variety of diagnostics ranging from the upper photosphere to the transition region \citep{PMJ_Jurcak_2010A, 2013ApJ...779..143R,PMJ_Tiwari_2016ApJ,Samanta2017, PMJ_Drews_2020A&A}.

These fine-scale bright dots are some of the smallest resolved heated structures in the lower solar corona. Investigating them further could provide key insights into the formation and evolution of larger coronal features. 

In this study, we analyze the formation of a plume in sunspot penumbra using the high resolution Extreme Ultraviolet Imager \citep[EUI;][]{EUI} on board Solar Orbiter \citep{SolO}, the Interface Region Imaging Spectrograph \citep[IRIS;][]{IRIS}, and the Atmospheric Imaging Assembly \citep[AIA;][]{AIA} and Helioseismic and Magnetic Imager  \citep[HMI;][]{HMI} both on board Solar Dynamics Observatory \citep[SDO;][]{SDO}. The formation of this plume coincides with the appearance of many small-scale bright dots at its base, and we investigate whether there is a relationship between EUV-bright dots and plume formation in sunspot penumbra. While there have been previous observations of existing plumes in sunspots, this is, to our knowledge, the first observation of the formation of a sunspot plume. This unique EUI-IRIS-SDO observation provides valuable insight into the mechanisms behind the formation of penumbral plumes.

In \autoref{sec:methods}, we introduce the data used and our alignment techniques. We discuss the region morphology, and characterize the plume's evolution (\autoref{sec:herecomestheplume}) and the bright dots at the plume base (\autoref{sec:dots}). We then analyze the dynamics of the region by creating space-time (\emph{x-t}) maps (\autoref{sec:x-t maps}) and by performing a wavelet analysis around the sunspot (\autoref{sec:wavelet}). Finally in \autoref{sec:conclusions}, we discuss our findings.

\begin{figure*}
    \centering
    \includegraphics[width=\textwidth-1.9cm]{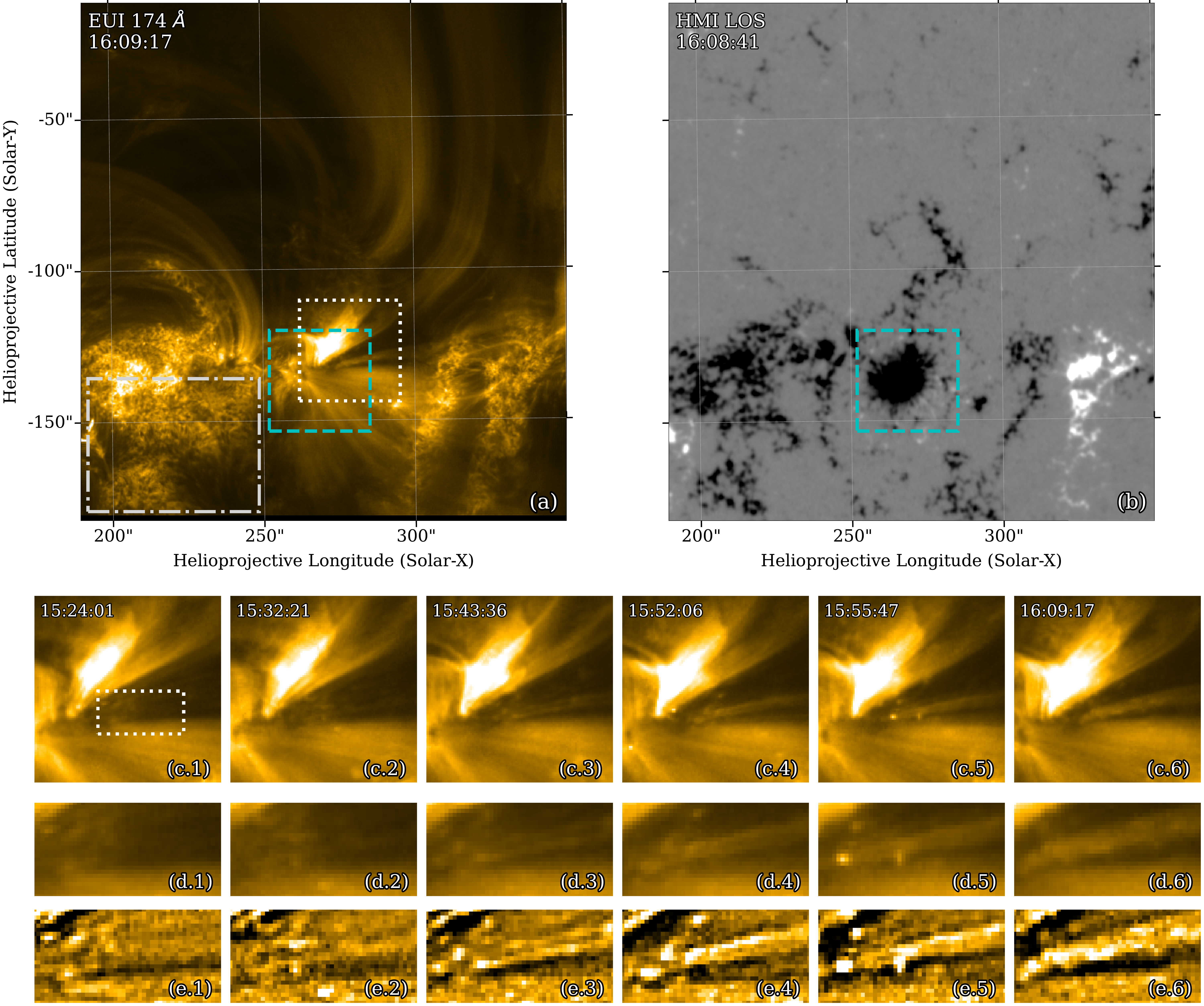}
    \caption{(a) Final frame in $\text{HRI}_{\text{EUV}}$ observation cut to the IRIS SJI field of view. The white dotted box shows the region shown in the (c) panels. The blue dashed box centers on the sunspot and is the region shown in Figure \ref{fig:contour}. The gray dashed-dotted box is the moss region used for aligning the $\text{HRI}_{\text{EUV}}$ observation with itself. (b) HMI line-of-sight magnetogram, where the blue dashed box is the region shown in Figure \ref{fig:contour}. (c.1)--(c.6) Zoom-ins of the region of interest over the course of the observation. These panels show the progression of plume formation, and times were chosen to highlight bright dots at its base. The white dotted box in (c.1) is the region shown in the (d) panels, which centers on the emerging plume. (e.1)--(e.6) $2\times2$ unsharp masked of the region shown in (d) panels. An animation of the zoomed-in region where the plume of interest develops is available in the online journal (see \autoref{sec:animation}).}
    \label{fig:context}
\end{figure*}

\section{Data and Methods}
\label{sec:methods}

We use calibrated L2 EUV data from the Extreme Ultraviolet High Resolution Imager ($\text{HRI}_{\text{EUV}}$) on board Solar Orbiter acquired on 2022 March 5 between 15:24:01 and 16:09:20 UT. This observational sequence focuses on a decaying sunspot near disk center (AR 12957, HARP \#8041, Figure \ref{fig:context}). During this observation, Solar Orbiter/EUI was situated 0.52 au from the Sun, which resulted in a final spatial scale of 185 km pixel$^{-1}$ at the solar surface. The spacecraft was well aligned with the Earth-Sun line (separation angle $\approx 5^\circ$). The $\text{HRI}_{\text{EUV}}$ wavelength passband is centered on 174 \AA{} and ranges from 171 to 178 \AA{} which is dominated by Fe IX (17.11 nm) and Fe X (17.45 and 17.72 nm) emission lines in the lower corona \citep{EUI}. The cadence of the $\text{HRI}_{\text{EUV}}$  images was 5 s. 

In order to connect the coronal structures with corresponding features in the transition region and chromosphere, we also use coordinated data from IRIS\footnote{\href{https://www.lmsal.com/hek/hcr?cmd=view-event\&event-id=ivo\%3A\%2F\%2Fsot.lmsal.com\%2FVOEvent\%23VOEvent_IRIS_20220305_150930_3600607418_2022-03-05T15\%3A09\%3A302022-03-05T15\%3A09\%3A30.xml}{IRIS observation}}.
We use slit-jaw imager (SJI) data in the 1400 \AA{} (Si IV) and 2796 \AA{} (Mg II k) channels. These datasets have a cadence of 10 s and a spatial scale of 240 km pixel$^{-1}$. Unfortunately, the IRIS spectrograph's raster did not cover the plume under study here (the short scan was centered on the sunspot), so we were not able to use the Mg II and Si IV spectral data.

We also use data from AIA and HMI. AIA is primarily used for alignment purposes but also to make direct comparisons with AIA's 171 \AA{} channel, which probes plasma at a similar temperature to EUI but at a $\sim 2$ times larger spatial scale (430 km pixel$^{-1}$) and lower temporal cadence (12 s). In order to obtain information about the region's magnetic configuration, we use HMI line-of-sight (LOS) magnetograms which have a cadence of 45 s and a spatial scale of 360 km pixel$^{-1}$. We also use HMI continuum images and inclination maps with a cadence of 12 minutes for context on umbra and penumbra locations, and for investigating the field properties.

\subsection{Preprocessing and Alignment}
\label{sec:alignment}
The data are processed, derotated, and aligned using SunPy routines \citep{sunpy_community2020}. 
After aligning all data using SunPy, we notice some residual errors in alignment, which we correct with cross-correlation techniques. To remove jitter in EUI images, we cross-correlate a subsection centered on a moss region that stays fairly consistent throughout the observation (gray dashed-dotted box in Figure \ref{fig:context}). 
After aligning EUI with itself, all the instrument pointings are relatively stable over this 45 minute observation window, so we calculate unique cross-correlation offsets for each corresponding image and apply the mean offset to all corresponding images. We cross-correlate the full field of view of similar wavelength channels among instruments to ensure all instruments are well aligned: EUI 174 \AA{} with AIA 171 \AA, IRIS 1400 \AA{} with AIA 1600 \AA, and HMI LOS with AIA 1600 \AA{}.
By adding the rms uncertainties of the EUI cross-correlation, EUI-AIA cross-correlation, and AIA-IRIS cross-correlation, we estimate the uncertainty in our alignment routine to be 430 km (1.8 IRIS pixels) in the horizontal direction and 215 km (0.9 IRIS pixels) in the vertical direction.

Due to the additional light travel time from Solar Orbiter's location to Earth, events appear $\sim 4$ minutes earlier in the EUI images than in those taken by instruments in geo- or sun-synchronous orbits (those situated 1 au from the Sun, e.g. IRIS, SDO). For ease of comparison, all the times shown in this paper are corrected to the detection time at 1 au.

In order to more easily compare intensity values across different instruments, we normalize each wavelength channel by its average intensity. For each frame in our time series, we compute the median intensity value of the field of view
(full region shown in Figure \ref{fig:context} panels (a) and (b)). Since the median intensity remains relatively constant throughout the observation in each channel (varying by at most 15\%), we take the mean value as our normalization factor. 
All figures and analyses in this paper use these normalized intensity values.

\begin{figure*}
    \centering
    \includegraphics[width=\textwidth-2.0cm]{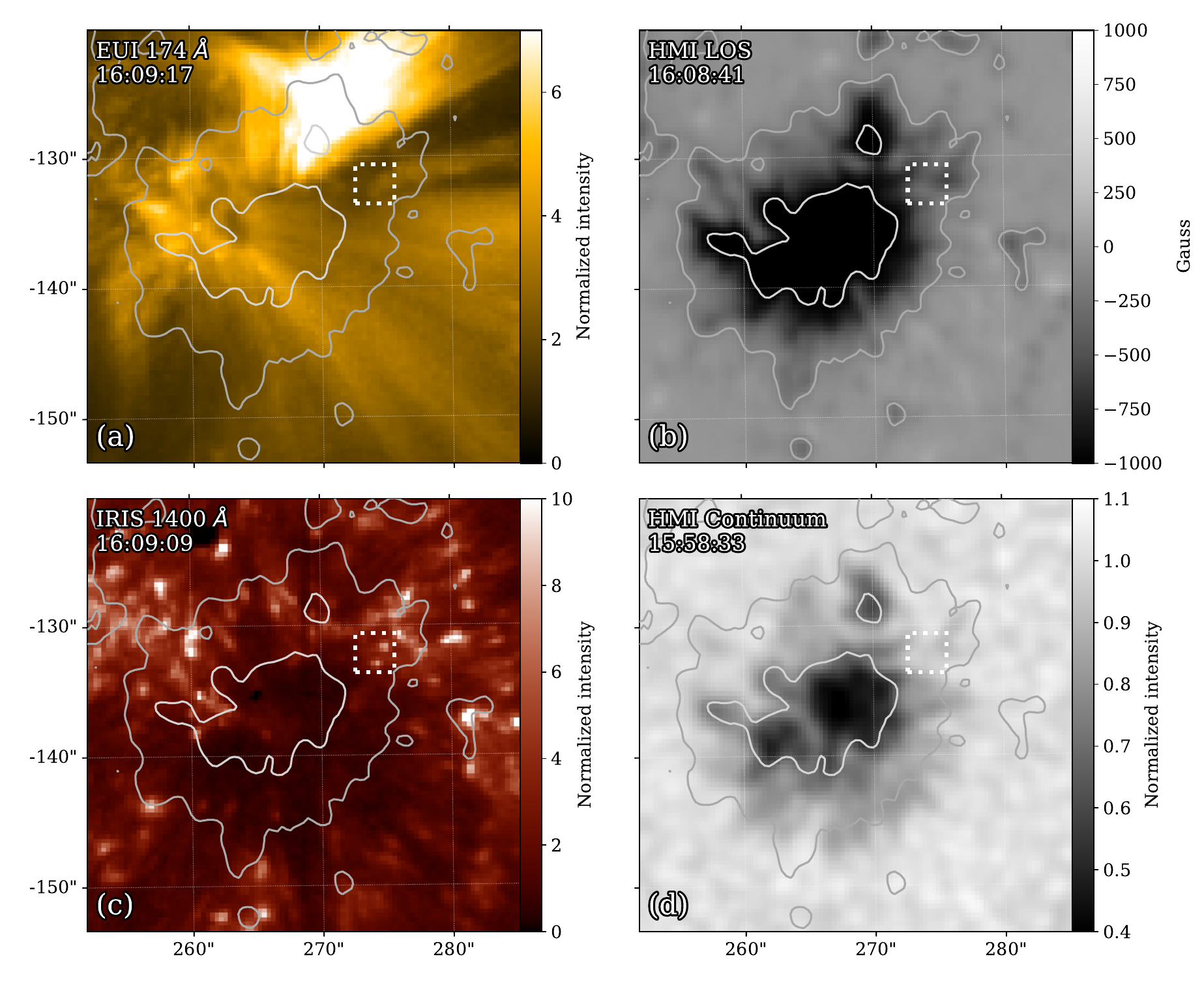}
    \caption{Zoom-in of the sunspot (see blue dashed box in Figure \ref{fig:context}). Panels (a) -- (d): EUI 174 \AA, HMI LOS magnetogram, IRIS 1400 \AA, and HMI continuum, respectively. Contours roughly outline sunspot umbra ($-1000$ G) and penumbra ($-150$ G). The white dotted box roughly indicates the position of the plume footpoint, which is in the penumbra.}
    \label{fig:contour}
\end{figure*}

\section{Region Morphology}
\label{sec:morphology}
\subsection{Plume Formation and Evolution}
\label{sec:herecomestheplume}
The most striking features in this coordinated observation are the formation of a plume and the high density of distinct bright dots at its base (Figure \ref{fig:context}). 
This plume forms as part of a dynamic system of active region loops that evolve rapidly.
At the beginning of the observation window, the 174 \AA{} map shows a gap, just west of the sunspot,  between two established, stable coronal structures, with the one to the north being significantly brighter and more compact (see Figure \ref{fig:context} (c.1)). At the end of the sequence, a long, collimated structure has developed in this gap (Figure \ref{fig:context} (c.6)); we define this feature as a plume due to its narrow appearance, and it having no obvious end-point as would be the case for a coronal loop. The plume structure is not visible in IRIS SJIs of 1400 \AA{} or 2796 \AA.

The magnetic configuration of this region consists of a sunspot within a large plage region (Figure \ref{fig:context} (b)). Both the sunspot and the closest plage section have the same negative polarity. The emerging plume seems to vanish above the negative-polarity plage just  west of the sunspot; it is not clear whether it continues above it, or whether it kinks upward to form a pseudostreamer. By looking at HMI data, we determine that the plume is rooted in the sunspot penumbra (Figure \ref{fig:contour}), in an area of a strongly inclined field. From the HMI inclination maps, the average inclination value of the photospheric field at the plume footpoint is $\sim120^{\circ}$ relative to the line of sight ($\sim30^\circ$ from the tangent to the solar surface, see \autoref{sec:hmi inclination}). This value does not vary significantly in this portion of the penumbra. 

\begin{figure*}
    \centering
    \includegraphics[width=\textwidth-2.8cm]{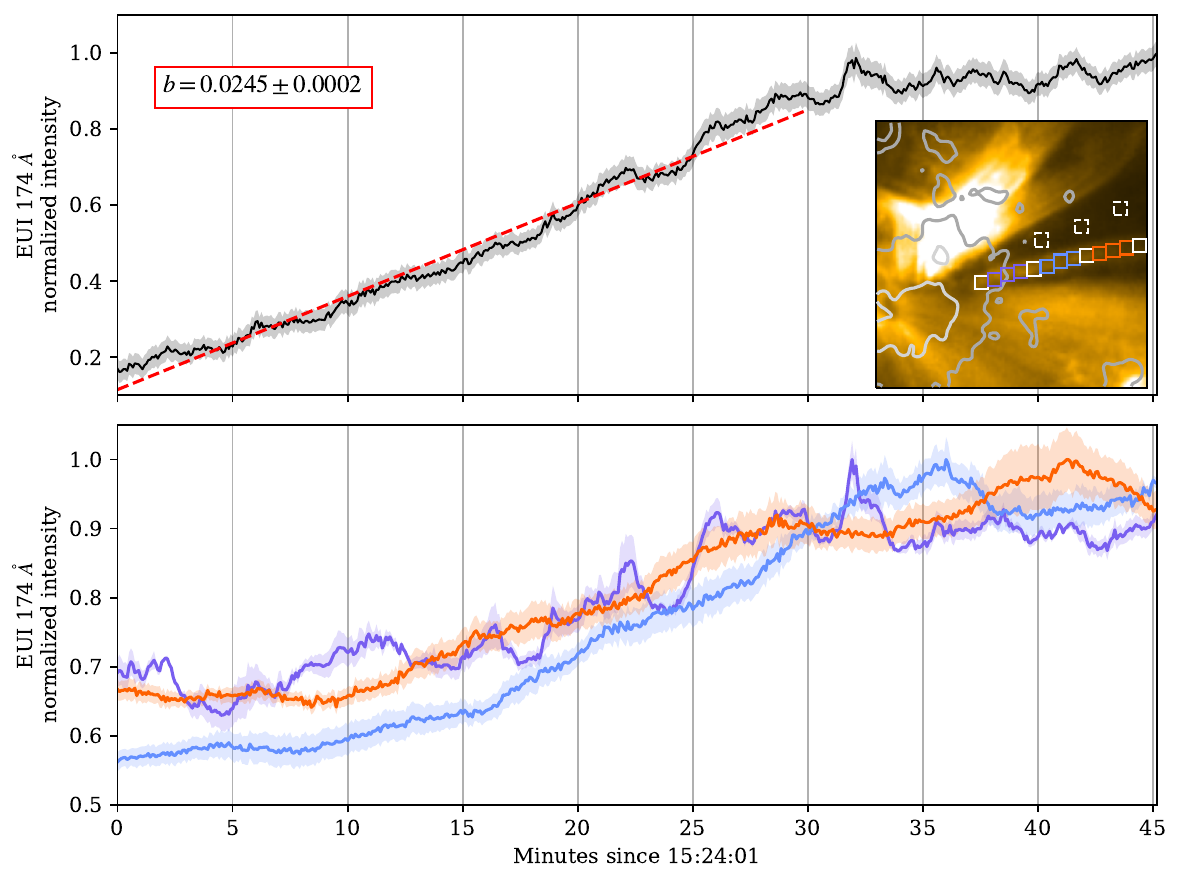}
    \caption{Normalized and background-subtracted EUI 174 \AA{} total intensity of the plume, over time. The inset shows EUI 174 \AA{} at the final frame of the EUI observation, and overlaid are the boxes we summed over to obtain the plume intensity (solid boxes, $5\times5$ pixels each), as well as the boxes used for background subtraction (white dashed boxes). The top panel shows the light curve for the entire plume (black curve, boxes $1-13$), and a linear fit to the rising intensity during the first 30 minutes of the observations (red dashed). The uncertainties (shaded regions) shown are derived from the standard deviation of the box intensities. The bottom panel shows the normalized light curves for three separate sections of the plume: the base (purple, boxes $2-4$), middle (blue, boxes $6-8$), and end (orange, boxes $10-12$). They all grow similarly in time, i.e., the plume increases in brightness at once over its full length.}

    \label{fig:plume_evo}
\end{figure*}

Starting at 15:24 UT, the overall intensity of the plume linearly increases in the EUI 174 \AA{} channel for approximately 30 minutes, at a rate of $\sim 2.5\%$ of the peak intensity per minute (Figure \ref{fig:plume_evo}, top Panel).
After 15:54 UT, the plume intensity plateaus at roughly 4 times its starting intensity, up until the end of the EUI observations.
The intensities we report are background subtracted using the average of three nearby boxes, which are also shown in Figure \ref{fig:plume_evo}. We subtract the specific background values at each time step in the series, but we check that the background intensity remains relatively constant during the observation interval (maximum change is $\sim 20\%$).

An important property of the developing plume is that the brightening of the structure appears to occur uniformly over its full length, over an interval of 30 minutes. Comparison of the light curves taken from different sections of the plume shows that there is no apparent time delay or outward propagation in the intensity enhancement (bottom panel of Figure \ref{fig:plume_evo}). It is hard to reconcile this behavior with what would be expected for mechanisms supposing the impulsive injection, and subsequent upward expansion, of hot material at the base of the plume. 
We estimate the visible length of the plume to be $\sim 13$ Mm in 174 \AA{} at its peak intensity and see no transverse oscillations during its lifetime.

While there is no EUI data of this region after 16:09 UT, we investigate this region in AIA 171 \AA{} and find that this plume lives until roughly 17:40 UT.
If we use the start of the EUI observation as the beginning of the plume's life, we obtain a plume lifetime of approximately 2.5 hr. 
Although plume lifetimes vary widely \citep[especially depending on the spatial scale being examined, see ][]{deforest_1997}, this is much shorter than the typical reported minimum plume lifetime of 20 hr \citep{Lamy_1997,Wang_2016,Avallone_2018}. However, it should be noted that selection effects in identifying stable plumes could skew the reported values toward longer lifetimes.

\subsection{Dot Detection and Analysis}
\label{sec:dots}

A notable feature in this region is the presence of several compact and dynamic brightenings. These ``dots'' are most visible in EUI 174 \AA{} and IRIS 1400 \AA{} and are present in many areas around the sunspot penumbra, but the base of the plume is the region where the dots are highest in number density, and are the most distinct and striking. They appear very compact and mostly circular, with no indications of directed, jetlike flows. 
Some of the dots exhibit proper motions, with excursions of several megameters, which are primarily aligned along the axis of the plume itself, indicating a possible connection between the two phenomena.

In order to understand the properties (size, lifetime, etc.) of these dots,
we manually track them.
For each dot detected in EUI 174 \AA{}, we take the same pixel position in IRIS 1400 \AA{} to determine if there is a corresponding dot lower in the solar atmosphere. To improve dot visibility for detection, we perform $2\times2$ pixel unsharp masking on EUI images (Figure \ref{fig:context} (e)). Similar to \cite{Tiwari_2022}, we use the following criteria to identify dots: (1) We select a quiet region and take the mean intensity and standard deviation---this is the $1\sigma$ noise level, and we only consider dots that are local maxima with an intensity of over $2 \sigma$, and (2) a dot must be detected in 2 or more frames, to ensure that detected dots are not noise.

\subsubsection{Dot Characteristics}
\label{sec:characteristics}
We use the following four parameters for characterizing dots: lifetime, speed, size, and intensity enhancement. Positional parameters---lifetime and speed---are derived from the unsharp masked images used for detection. Parameters dependent on intensity---size and intensity enhancement---use nonenhanced images.
These are calculated as follows:
\begin{itemize}
    \item \emph{Lifetime.} Subtract the last time the dot is visible from the first time the dot is visible.

    \item \emph{Speed.} To calculate the dot's plane-of-sky maximum velocity, we find the dot's maximum displacement ($\delta d$), and the time difference of this maximum displacement ($\delta t$). The speed of each dot is $\delta d / \delta t$. 
    
    \item \emph{Size}. We fit a 2D Gaussian when the dot is at its peak intensity. Because our dots are roughly circular, we calculate the full width at half-maximum (FWHM) along their major and minor axes and report the mean value as the dot size.
    
    \item \emph{Intensity enhancement}. We take vertical and horizontal cross sections centered on the dot's peak pixel. To determine the background, we use pixels located at a distance of the dot's radius plus 3 pixels. The average intensity of these four background pixels is taken as the background value. The intensity enhancement is calculated by dividing the dot's peak intensity by the background value and multiplying the result by 100. We report the intensity enhancement when the dot is at its maximum intensity.
\end{itemize}

Over the course of the observation, we detected \NumDots{} dots at the plume base in EUI 174 \AA. The mean parameters of interest for all dots detected and their IRIS 1400 \AA{} counterparts are shown in Table \ref{table:params}. Since we only track dots in EUI 174 \AA{}, we report only the 174 \AA{} lifetime and velocity.
\begin{deluxetable}{l | c c c }[H]
    \tablehead{\colhead{Parameter} & \colhead{EUI 174 \AA} & \colhead{IRIS 1400 \AA}}  
    \startdata
    Lifetime (s) & \DotLifetime{} $\pm$ \DotLifetimeStd{} & -- \\
    Max speed (km s$^{-1}$) & \DotSpeed{} $\pm$ \DotSpeedStd{} & -- \\
    Size (km) & \EUIsize{} $\pm$ \EUIsizeStd & \IRISsize{} $\pm$ \IRISsizeStd \\
    Intensity ($\%$) & \EUIintensity{} $\pm$ \EUIintensityStd & \IRISintensity{} $\pm$ \IRISintensityStd
    \enddata
    \caption{Mean values of parameters of interest for EUI 174 \AA{} dots and the corresponding IRIS 1400 \AA{} dots. The uncertainties reported are the standard deviations of the mean.}
    \label{table:params}
\end{deluxetable}

There is no significant size difference between dots in EUI 174 \AA{} and IRIS 1400 \AA. Dots are significantly brighter with respect to their environment in the IRIS 1400 \AA{} passband than in EUI 174 \AA{}, and we find a positive correlation between peak intensity in the two channels, i.e. a brighter dot in 174 \AA{} corresponds with a brighter dot in 1400 \AA.
We find no consistent timing offset in when a dot reaches its peak intensity in EUI 174 \AA{} versus IRIS 1400 \AA{}.

The average lifetime of the dots we detect align with previous observations of dots in emerging flux regions \citep[$\sim 50$s,][]{Tiwari_2022}, which appear to originate from magnetic reconnection, as well as dots in the transition region in sunspots \citep[$\sim 40$s,][]{Tian_2014}, which originate either from magnetic reconnection or strong downflows.
The dot properties also align with dotlike campfires \citep{Berghmans_2021, Panesar2021}, and fine-scale dotlike heating events seen by Hi-C 2.1 \citep{Tiwari_2019}, all using wavelength passbands centered on Fe IX/X emission.

\section{Temporal Analysis}
\subsection{Space-time (\emph{x-t}) maps}
\label{sec:x-t maps}
\begin{figure*}
        \centering
        \includegraphics[width=\textwidth-1.2cm]{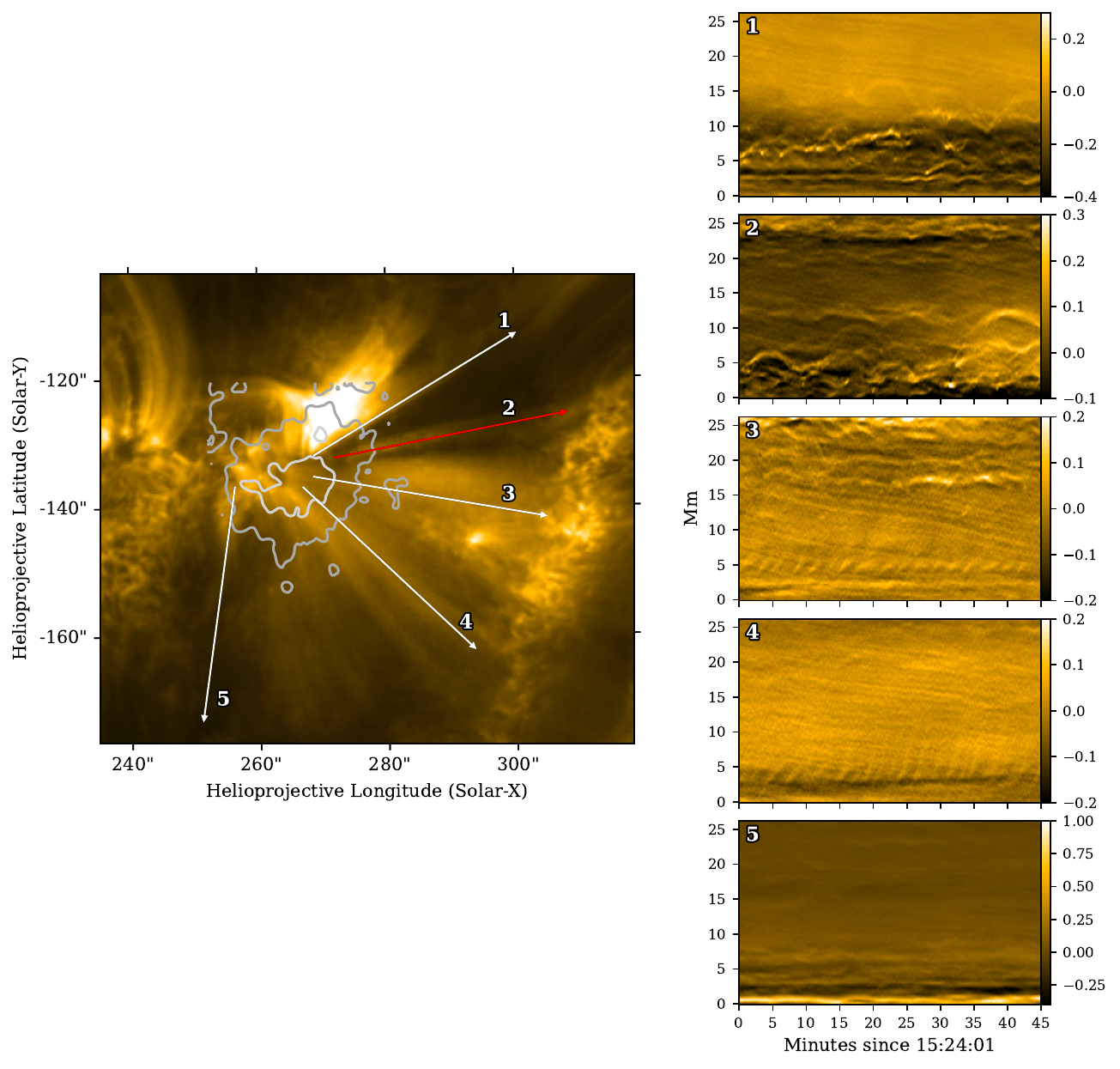}
        \caption{EUI 174 \AA{} \emph{x-t} maps for various slices around the sunspot. To indicate the location of the umbra and penumbra, we show the contours as plotted in Figure \ref{fig:contour}.
        Slices correspond to the following features: a bright coronal feature in the northwest of the region (slice 1); the plume studied in this work (slice 2, red); portions of the southern loop (slices 3 and 4), a diffuse coronal feature (slice 5). Slices are produced from $2\times2$ unsharp masked images, and averaged over a width of 6 pixels. All slices are about 25 Mm long, and the arrows indicate the direction of the spatial axis in the \emph{x-t} maps, with zero position nearest to the sunspot. Note that the tops of both slices 2 and 3 overlap with the plage region west of the sunspot.
        }
        \label{fig:multiple cuts}
\end{figure*}

\begin{figure*}
    \centering
    \includegraphics[width=\textwidth-1.3cm]{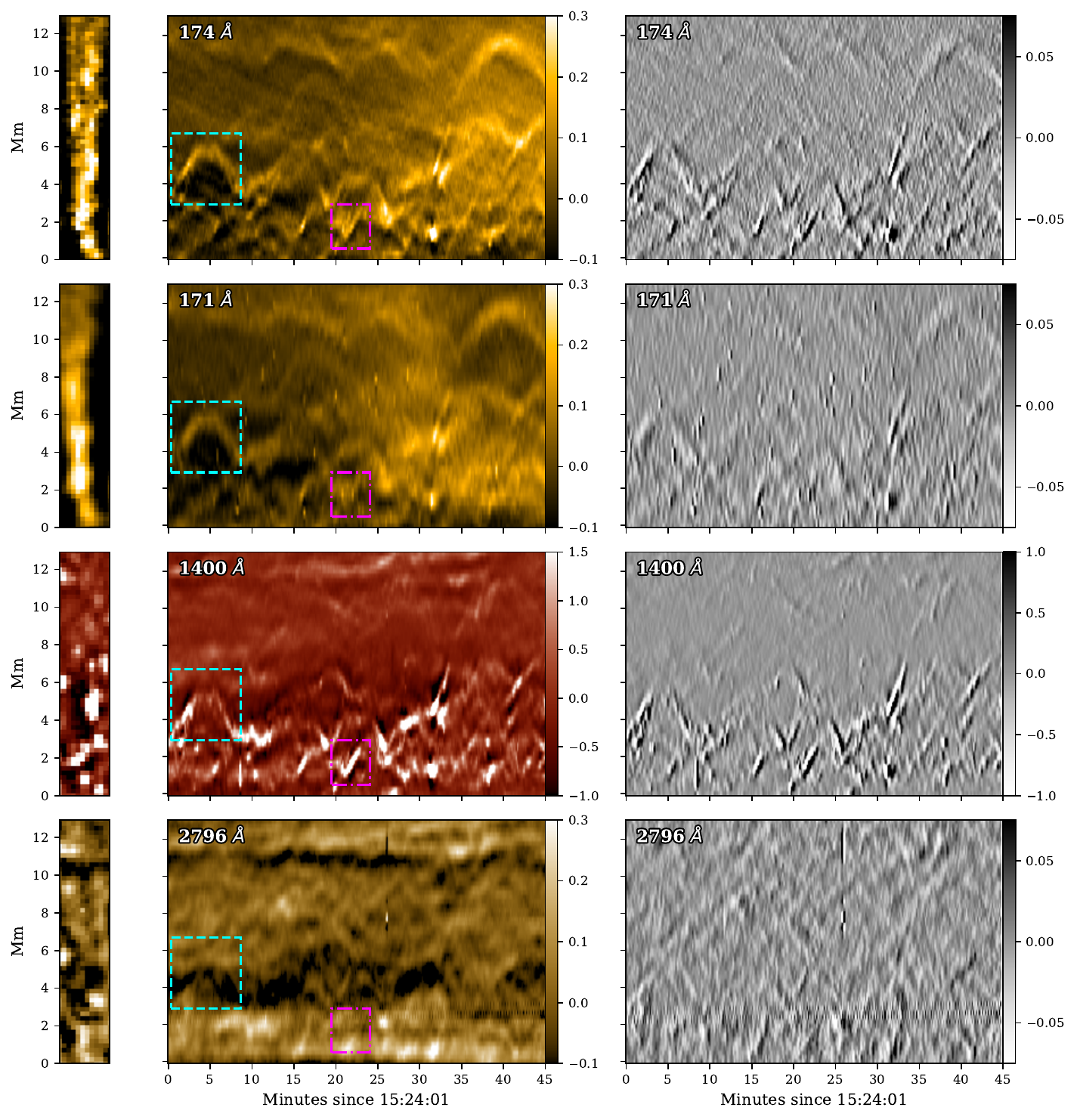}
    \caption{\emph{x-t} maps along the plume in various wavelengths (zoom-in of slice 2 in Figure \ref{fig:multiple cuts}). The first column shows our slit location at plume maximum brightness (the final image in the observation). The running difference images shown in the third column are made with a time step of 15 s. The cyan dashed box and the pink dashed-dotted box identify a clear parabolic feature and "V"-shaped feature, respectively.}
    \label{fig:x-t maps}
\end{figure*}

By tracking the bright dots described in Section \ref{sec:morphology}, we find that a subset of them exhibit a preferential movement aligned with the plume axis.
To more clearly reveal their dynamics, we create \emph{x-t} maps. We define an artificial slit with a width of 6 pixels along the length of the plume and plot the average intensity over these pixels along the slit as a function of time, for the whole duration of the EUI observations. For comparison, we do the same for four other radial slices around the sunspot, tracing different coronal regions. Figure \ref{fig:multiple cuts} shows the resulting plots. The maps have been created using the  
$2 \times 2$ unsharp masked EUI 174 \AA{} images.

The \emph{x-t} maps reveal very different temporal behaviors among the five slices.
The southernmost slice, labeled 5, does not lie along any distinct coronal structure and does not display obvious temporal intensity patterns. 
On the contrary, slices 3 and 4 lie along an established and relatively static loop, and display a clear pattern of periodic slanted ridges, originating within or near the sunspot umbra, and extending for several Mm along the loop. Without detailed spectral information, we lack the ability to specifically assess the drivers of this pattern, but given their periodicity of about 2–3 minutes, these ridges are most likely manifestations of slow magnetoacoustic waves originating in the sunspot and propagating along field lines \citep{Berghmans_1999,De_Moortel_2002, Skirvin_2024, Zhao_2025}.
The faint downward stripes seen in slices 1, 2, 3, and 4 (speeds $\sim 7$ Mm h$^{-1}$) are most likely small flat-field artifacts that appear to move eastward after correcting for solar rotation (D. Berghmans 2025, private communication).

The most striking temporal pattern is observed along the emerging plume (slice 2), where multiple parabolic tracks are clearly noticeable, with an apparent periodicity of $\sim 7-10$ minutes. The period of the parabolas seems to increase with increasing distance along the plume, but most of these patterns are seen within the first $\sim 10$ Mm from the plume's base. 
The bright, preexisting coronal structure in the north (slice 1) also shows some similar parabolic behavior near its base, but with smaller, less coherent dynamic motions.

Parabolic tracks similar to those observed along the plume have been reported most often in chromospheric diagnostics and interpreted as a signature of spicules, dynamic fibrils, or shock-related phenomena in general \citep{Hansteen_2006,De_Pontieu_2007_b,De_Pontieu_2007}. Similar patterns have also been observed at transition region and coronal signatures, such as in IRIS data \citep[e.g.][]{Skogsrud_2016} and, most recently, in EUI 174 \AA{} by \citet{Mandal_2023,mandal2023evolution}. The latter
authors highlight how the enhanced resolution of EUI clearly allows their identification as the hot tips of dynamic fibrils, reaching temperatures of $1-1.5$ MK. 

We can obtain further insight into the EUI dynamics along the plume by comparing the multiple wavelengths in our dataset. Figure \ref{fig:x-t maps} displays the EUI \emph{x-t} map for slice 2, together with the  corresponding slices in AIA 171 \AA{}, IRIS 1400 \AA{}, and IRIS 2796 \AA{}. The panels in the last column display running difference images to better highlight spatiotemporal patterns, and the length of the slice is limited to the initial $\sim 12$ Mm for better clarity.

Multiple parabolic tracks are well discernible in all wavelengths (see e.g. the parabola within the cyan square in Figure \ref{fig:x-t maps}), although the EUI map provides the clearest picture. Most EUI 174 \AA\ features can also be seen in the AIA 171 \AA\ maps, especially once identified in the higher-resolution, higher contrast EUI images. 
Similar to that reported in \cite{mandal2023evolution}, we also identify many of the same tracks as bright features in IRIS SJI 1400 \AA\ (especially when analyzing the running difference images) and, to a lesser extent, as darker features in the IRIS SJI 2796 \AA\ images. While there does not seem to be an obvious time delay between the appearance of a feature at any given wavelength, we note that the 2796 \AA\ map shows parabolas extending to shorter distances along the plume.

Finally, fits to multiple parabolic tracks show that these features have an approximate acceleration of $-0.02$ km s$^{-2}$ and an initial velocity of 12 km s$^{-1}$. Both these values are comparable to the low end of their respective distributions, as found in \cite{Pereira_2012} for type I spicules, and in \citet{Mandal_2023} for ``moss'' dynamic fibrils. All these observational findings confirm the scenario of dynamic fibrils developing along the plume, with the transition region and coronal signatures representing the hot tips of the chromospheric counterpart.

Alongside parabolic features, in some locations we observe V-shaped features (e.g., in the pink dashed–dotted box in Figure \ref{fig:x-t maps}) --- most strikingly in 174 \AA{} and 1400 \AA{}.
To our knowledge, such V-shaped features have not been reported in EUI or IRIS observations previously, but they appear to occur when the end of a dynamic fibril event coincides with the beginning of the next one recurring in the same location. Given that the fibrils appear brighter (hotter) during their ascending phase, and toward the end of their evolution (as clearly seen for the parabola framed by the cyan box in Figure \ref{fig:x-t maps}), this combination of events can account for the observed bright V shapes.
This behavior is most obvious in 1400 \AA, and aligns with the findings of \cite{Skogsrud_2016}, that bright grains have the strongest emission in 1400 \AA{} on the upward ascent phase of the dynamic fibril. \citet{mandal2023evolution} also report similar intensity evolution for fibrils observed in EUI 174 \AA. 

Due to the lack of spectroscopic information, we cannot obtain conclusive information about the presence and amplitude of inflows/downflows along the plume. Yet, the structure appears very inclined from the local normal, as also inferred from the inclination of the photospheric magnetic field at its base (see \autoref{sec:hmi inclination}) and from the long periodicities observed in the dynamic fibrils along its length (see discussion in Section \ref{sec:conclusions}). Hence, any significant plasma flow along the structure should be fairly visible in the (plane-of-the-sky) \emph{x-t} maps of Figures \ref{fig:multiple cuts} and \ref{fig:x-t maps}, but we do not observe such a pattern at any time during the plume's evolution. This conflicts with the idea that inflows are a necessary component of sunspot plumes \citep{Brynildsen2001}.

\subsection{Wavelet analysis}
\label{sec:wavelet}

\begin{figure*}
    \centering
    \includegraphics[width=\textwidth-2cm]{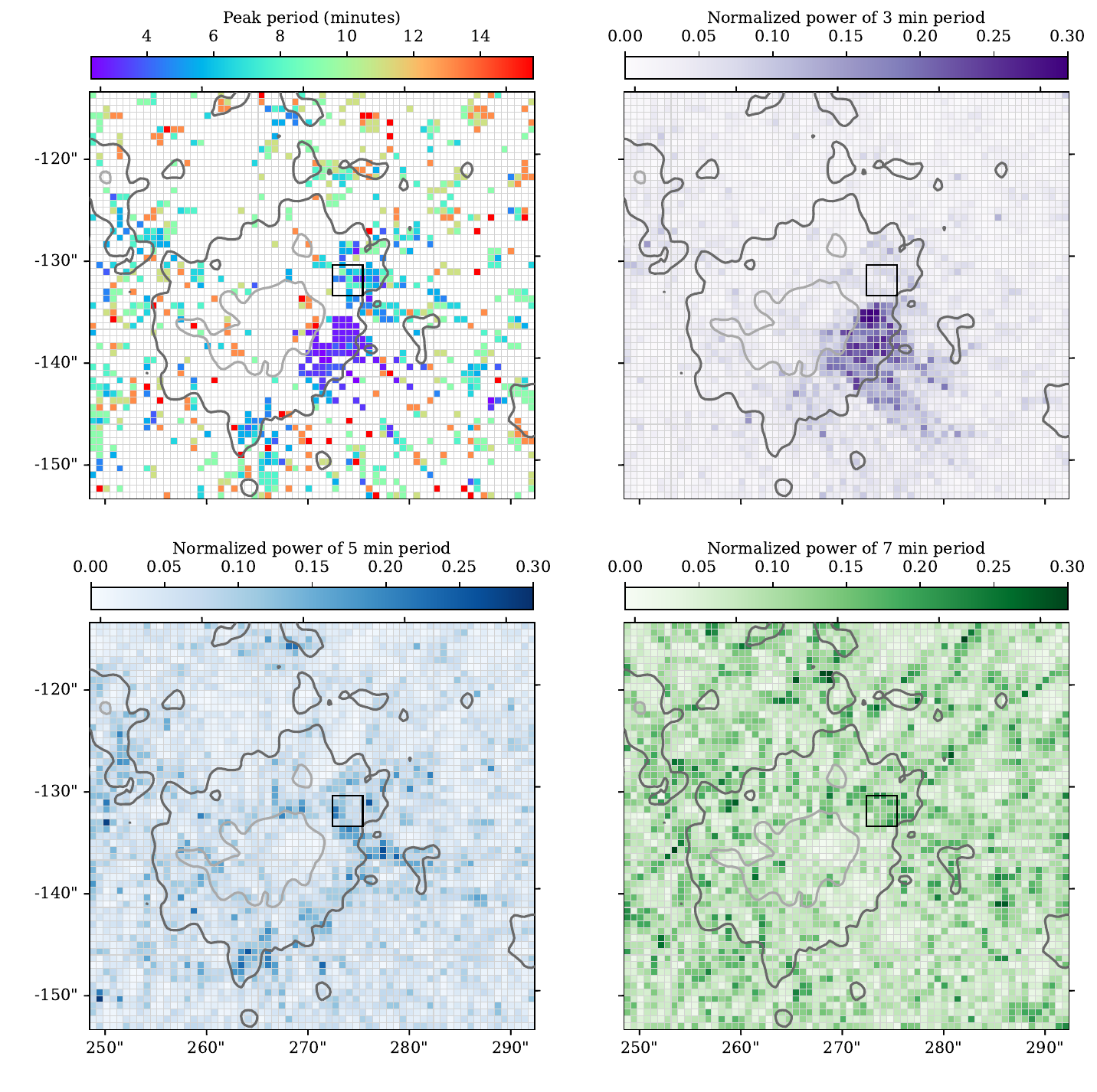}
    \caption{Wavelet analysis of the EUI 174 \AA{} time series. The top left panel shows the period with the maximum power. The positions filled with white are regions where the peak period lies outside the wavelet analysis cone of influence. The other three panels show the normalized power of 3, 5, and 7 minute periods. The plume footpoint is outlined by the black box, and the contours roughly outline sunspot umbra and penumbra (same locations as in Figure \ref{fig:contour}).}
    \label{fig:wavelet}
\end{figure*}

Previous analyses of coronal plumes have suggested that magnetic reconnection at plume bases could provide the required energy and mass via ejecta of heated material, often referred to as “jetlets” \citep{Raouafi_2014,Avallone_2018,Panesar_2018}. 
In particular, small-scale jets may occur with a periodicity of $\sim 3-5$ minutes, which hints that the global \emph{p}-mode waves are an important driver of these features \citep{Kumar_2022}.
While we observe no jetlets in our data, the \emph{x-t} map analysis suggests that at least some of the brightenings at the plume base present an oscillatory behavior, potentially related to the global \emph{p}-modes. 

To better investigate possible periodicities in the plume's area, we perform a Morlet wavelet analysis \citep{Torrence_1998} of the EUI 174 \AA{} intensity time series (a Fourier transform analysis provided very similar results). We divide the region around the sunspot into an evenly spaced grid of $2\times2$ pixel regions, and create the intensity time series for each grid point. We detrend the individual time series by fitting third-degree polynomials, and finally smooth the data by linearly convolving with a kernel of size 5 time steps before performing the wavelet analysis. The results are shown in Figure \ref{fig:wavelet}. White grid points are regions where a reliable value for the periodicity could not be determined. 

The top left panel of Figure \ref{fig:wavelet} shows the periodicity at which maximum power is found in any given grid position. Within the sunspot and its surroundings, only two clearly coherent patches of power at a given periodicity are observed. The first one, at roughly ($275\arcsec$, $-140\arcsec$), is the source region of the loops from which slices 3 and 4 of Figure \ref{fig:multiple cuts} are drawn, and hosts maximum power at $\sim 2-3$ minutes periodicity. This is consistent with the pattern of slanted ridges observed in the \emph{x-t} slices.
The second one, about 10$\arcsec$ north of the first patch, encompasses the footpoint of the plume --- the small area framed by the square black box in the Figure, and displays much longer periodicity, around 7–10 minutes. This is again consistent with the pattern observed in the \emph{x-t} maps for slice 2 (and, to a lesser extent, for slice 1) in Figure \ref{fig:multiple cuts}.

The other three panels of Figure \ref{fig:wavelet} show maps of the power at selected periodicities, created by integrating the power over a 1 minute range around the central value, and normalizing the result to the total wavelet power in the $1-15$ minutes range (i.e. a value of 0.05 corresponds to that period being responsible for 5$\%$ of that grid position's total global wavelet power).
Oscillations around 3 minutes represent the dominant power only for a part of the loop system (top right panel), while 5 minutes' power is most relevant along the western edge of the penumbra (bottom  left panel). Periodicities of 7 minutes or longer provide the most power in many small-scale areas, scattered throughout the region,
but seem to be most spatially coherent at the base of the plume (bottom right panel).

We note the absence of 3 minute power within the umbra in our observations. While previous studies have reported umbral oscillations in the AIA 171 \AA{} filter \citep{Reznikova2012, Chai2022}, there has been no systematic study of how frequent such oscillatory signatures are across a range of sunspots, especially for small sunspots like the one in our data. \cite{Wu2023}, studying two different sunspots, saw differences in the oscillatory behavior at coronal heights that they attributed to the open versus closed magnetic topology of the features.

The spatial distribution of dominant periodicities appears to be linked to the inclination of the magnetic field relative to the local vertical along which the coronal structures form. The cutoff frequency for the propagation of (magneto-)acoustic waves originating at the solar surface shifts to lower values when the magnetic field is inclined \citep{De_Pontieu2004, DePontieu2005, 2023LRSP...20....1J}. 

Figure \ref{fig:mag inclination} in \autoref{sec:hmi inclination} shows the HMI map of field inclination for our region, over the same FOV as Figure \ref{fig:wavelet} (as stated above, the map is derived from the combination of HMI data over a period of $\sim 12$ minutes).
The coronal loops traced by slices 3 and 4 in Figure \ref{fig:multiple cuts} appear to be rooted near the interface between the umbra and penumbra, and the observed periodicities reflect the dominant, chromospheric 3 minute peak \citep{Schrijver1999}. 
On the other hand, the newly formed plume appears to be originating from a location in the penumbra with a large inclination --- about $\sim 120^\circ$ on average, which would allow the propagation of (magneto-)acoustic waves at longer periods (scaling with the cosine of the inclination angle).

\section{Discussion and Conclusions}
\label{sec:conclusions}

We studied the formation of a narrow coronal plume rooted in sunspot penumbra using a unique set of coordinated observations from Solar Orbiter/EUI, SDO/AIA, and IRIS. The improved spatial resolution of EUI, more than twice that of AIA, allows us to better resolve and identify some of the dynamical behavior that occurs during the plume formation. 

Our EUI time series shows the onset of a plume apparently rooted in the penumbra of the leading sunspot of AR 12957. During our observing sequence, the EUV intensity of the plume increases linearly by $\sim 2.5\%$ of its peak intensity per minute for the first 30 minutes of the observation, and then plateaus at roughly 4 times its starting intensity. Notably, the plume appears to brighten coherently over its full length rather than showing evidence of a progressive upward extension. 
This is a new observational result that provides a strong constraint that should be taken into account in models of plume formation.

The enhancement of this plume corresponds with a set of striking circular dots at its base. We used a manual detection method to characterize the bright dots, finding that they have an average lifetime of \DotLifetime{} s, a maximum velocity of \DotSpeed{} km s$^{-1}$, a size of \EUIsize{} km, and an intensity enhancement of \EUIintensity{}$\%$ in EUI 174 \AA. The corresponding dots in IRIS 1400 \AA{} are typically brighter (average intensity enhancement of \IRISintensity{}$\%$), but otherwise have similar characteristics.
Unlike previous studies of bright dots in sunspot penumbrae \citep{Kleint_2014,Tian_2014,Alpert_2016}, we see little to no jetlike extensions from the features in this dataset.
The correlation between size, intensity, and timing of peak intensity in these two channels suggests that the dots may be multithermal or we are seeing transition region emission in the EUI 174 \AA{} passband. The EUI 174 \AA{} passband does include O V and O VI lines, which form in the TR at about 300,000 K, so features cooler than 1 MK may have been captured \citep{Tiwari_2022}.

We find that a subset of these bright dots preferentially move along the plume axis. To further investigate these dynamics, we created \emph{x-t} maps and found $\sim 7-10$ minute parabolic patterns at multiple locations along the length of the plume. Such long periodicities have been associated with highly inclined magnetic fields \citep{Lites1988}. The region of the plume is not the only location around the sunspot with such features, but it is important to note that this is where the periods are longest, as confirmed by a wavelet analysis, and the tracks are most distinct. The oscillatory motions of the dots are present throughout the 45 minute time series.
We note that the significantly shorter average dot lifetime compared to the lifetime of parabolic features at the plume base can be partly attributed to our dot detection method, which imposes an intensity threshold of $2\sigma$ above the background intensity.
A single parabolic feature may be identified as multiple dots due to intensity variations causing intermittent detection above and below the threshold.

The repeated, parabolic motions of the brightenings seen in the \emph{x-t} maps suggest an upward acceleration of material that slows due to gravity and falls back downward. This behavior is distinct from the slanted ridges appearing in other regions around the sunspot, which have been interpreted as outwardly propagating magnetoacoustic waves \citep{Berghmans_1999,Prasad_2012,Mandal_2023}. Instead, these features appear to be similar to the signature of magetoacoustic shocks that dissipate energy in the chromosphere and above, and create ballistic trajectories of matter moving into the corona. Recently, similar parabolic patterns have been reported in EUI HRI data \citep{Mandal_2023,mandal2023evolution}. The bright dots we observed in IRIS 1400 \AA{} and EUI 174 \AA{} are likely the shock wave compression on the top of chromospheric dynamic fibrils \citep{Hansteen_2006, De_Pontieu_2007}. 
Some previous works have shown, theoretically and observationally, that shock-driven jets can lead to the heating of material to coronal temperatures \citep{DePontieu_2017,Sykora_2017}. For this study, we lack the spectroscopic and chromospheric observations that would allow us to confirm that our dots are the result of a similar process. Future, multiviewpoint observations might help determine how the parabolic features could be related to the extended, continuous formation of the plume.

There are other mechanisms that have been proposed as fundamental drivers of plume formation in regions other than sunspots (in coronal holes or quiet solar regions), but none of which seem to be present in the plume we have analyzed.
The apparent occurrence of magnetic reconnection at the base of plumes has been associated with jetlets, directed flows of material along field lines extending upward away from the reconnection site \citep{Raouafi_2014,Malaker_2024}. We do not see any such extensions coming from the bright dots, which are essentially circular even at the high spatial resolution that was achieved in these observations.
Mixed-polarity magnetic field in sunspot penumbra \citep{Tiwari_2015} could create sites for such processes to occur. We only have the limited spatial resolution of the magnetic field maps obtained with SDO/HMI for this dataset, which do not resolve any small-scale mixed-polarity fields that could be the source of such reconnection (see \autoref{sec:hmi res}). Higher spatial resolution magnetic maps would be needed to probe this question, with direct measurements of the chromospheric field from the Daniel K. Inouye Solar Telescope \citep[DKIST;][]{DKIST2020} providing valuable insights into the formation mechanism(s) of these dots.

Alternatively, previous studies have reported downflowing transition region and coronal plasma as the cause for umbral and penumbral brightenings  \citep{Kleint_2014, Deng_2016} and coronal plumes \citep{Brynildsen2001,Chen_2022,Moore_2023}. However, we find no evidence of strong, consistent downflows along the plume under investigation, and coronal condensations \citep{Li_2021} do not appear to be consistent with the initial upward/outward motions seen in the region around this plume. 
Spectroscopic information, from IRIS or the upcoming Multi-Slit Solar Explorer \citep[MUSE;][]{MUSE2020}, would help distinguish among these three mechanisms by characterizing the direction and Doppler velocity of any apparent motions.

The presence of these different mechanisms manifesting themselves, often in isolation, in a variety of plume observations, is an indicator that perhaps there is not a single, essential mechanism that drives all plume-formation scenarios. Each of these processes may work optimally in specific physical conditions, in particular, the topology and complexity of the magnetic field in which the plume appears. The appearance of plumes may not be sufficient evidence to assume the presence of specific processes or magnetic configurations.

The parabolic motions of the bright dots are suggestive of a relationship to dynamic fibrils, similar to the connection between plumes and spicules that has already been proposed \citep{Wilhelm_2000, Cho_2023}. However, their physical connection to the actual formation, heating, or mass loading of the plume is hard to establish, given that the plume itself appears like the sum of multiple optically thin components along the line of sight.
The parabolic motions we observe are present, with equal strength and comparable periods, at multiple locations along the plume (over a length of $\sim 12$ Mm). For a simple, idealized coronal loop, the injection of energy by the magnetoacoustic shocks would occur only close to where that loop was rooted in the penumbra and not further along the loop where it is already at coronal heights. Their visual alignment in this case could be due to a common orientation of the field at multiple heights in this region, with the plume itself having multiple footpoints. The optically thin contributions from that set of field lines could overlap vertically to produce the appearance of a single structure.

Although distinguishing between correlation and true physical driving is a general challenge in plume studies, our analysis suggests that jetlets and downflows may not be necessary components of plume formation. Instead, plume-formation mechanisms in the sunspot penumbra may be influenced by dynamic fibrils.

\section*{Acknowledgments}
We would like to acknowledge the valuable discussions we had with participants of the 11th Coronal Loops Workshop. We would also like to thank Navdeep Panesar for advice on retrieving SDO data and David Berghmans for discussions about Solar Orbiter/EUI alignment. We thank the referee for their constructive comments, which have improved the clarity and content of the paper.

Support for this study was provided by IRIS, and Ayla Weitz acknowledges the support of the University of Colorado’s George Ellery Hale Graduate Student Fellowship.
S.K.T. gratefully acknowledges support by NASA HGI (80NSSC21K0520) and HSR (80NSSC23K0093) grants, and NSF AAG award (no. 2307505).
K.P.R. acknowledges support by NASA grants 80NSSC22K0712 and  80NSSC20K1282.
The National Solar Observatory is operated by the Association of Universities for Research in Astronomy, Inc. (AURA), under cooperative agreement AST-1400450 with the US National Science Foundation.
Solar Orbiter is a space mission of international collaboration between ESA and NASA, operated by ESA. The EUI instrument was built by CSL, IAS, MPS, MSSL/UCL, PMOD/WRC, ROB, and LCF/IO. IRIS is a NASA small explorer mission developed and operated by LMSAL with mission operations executed at NASA Ames Research Center and major contributions to downlink communications funded by ESA and the Norwegian Space Centre. The AIA and HMI data are courtesy of NASA/SDO and the AIA and HMI science teams.

\pagebreak

\appendix
\twocolumngrid

\section{Magnetic Field Inclination}
\label{sec:hmi inclination}
\begin{figure}[H]
    \centering
    \includegraphics[width=\linewidth]{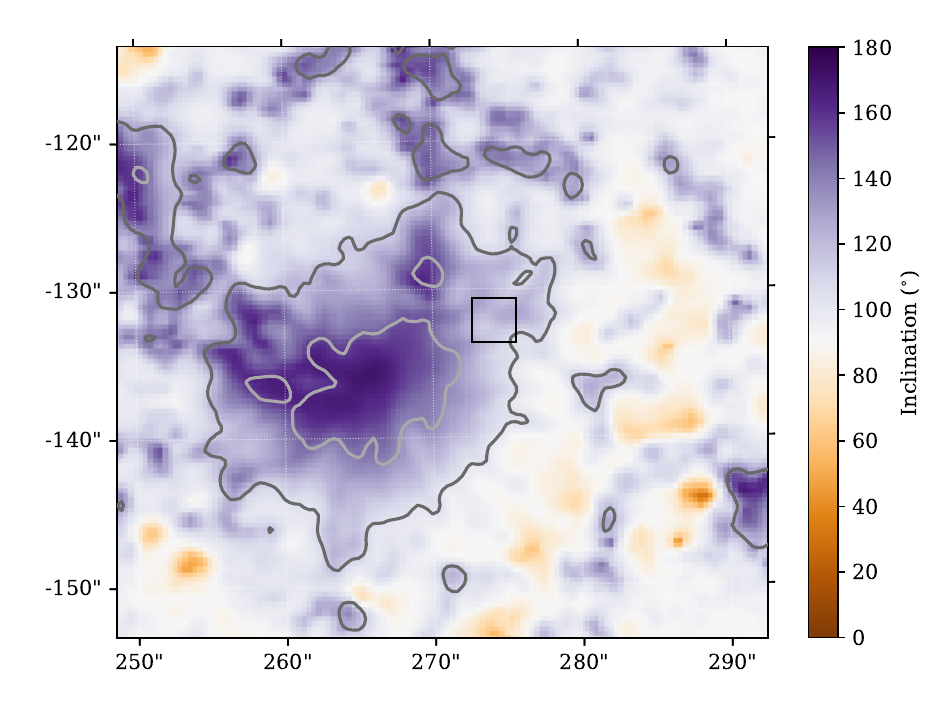}
    \caption{HMI inclination map at 15:58 UT (same field of view as Figure \ref{fig:wavelet}). The plume footpoint is outlined by the black box, and the contours roughly outline sunspot umbra and penumbra (same locations as in Figure \ref{fig:contour} but at a different time). White indicates completely horizontal field ($90^{\circ}$), while purple and orange indicate negative and positive polarity vertical fields.}
    \label{fig:mag inclination}
\end{figure}

The mean inclination of the photospheric magnetic field at the plume footpoint is $\sim 120^{\circ}$, which is not significantly different from regions elsewhere around the sunspot (Figure \ref{fig:mag inclination}).

\vfill\null
\section{Change in Magnetic Flux at Plume Base}
\label{sec:hmi res}
\begin{figure}[H]
    \centering
    \includegraphics[width=\linewidth-1.0cm]{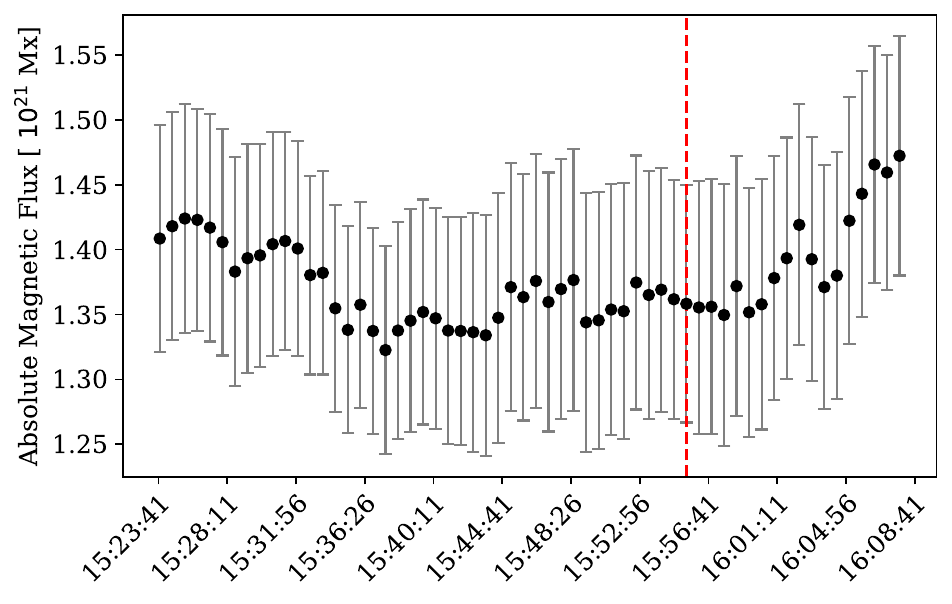}
    \caption{Absolute magnetic flux of the plume footpoint throughout the observation (white dotted box in Figure \ref{fig:contour}). The red dotted line shows the approximate time when the plume intensity plateaus. We estimate the uncertainty on our flux values by taking the rms of the difference between our footpoint and eight surrounding regions shifted by 1 pixel.}
    \label{fig:mag flux}
\end{figure}

At the plume footpoint, we sum the absolute magnetic flux derived from the HMI LOS magnetograms over the course of the observation. We find no significant change in absolute magnetic flux throughout the observation (Figure \ref{fig:mag flux}).

\begin{figure}[H]
    \centering
    \includegraphics[width=\linewidth]{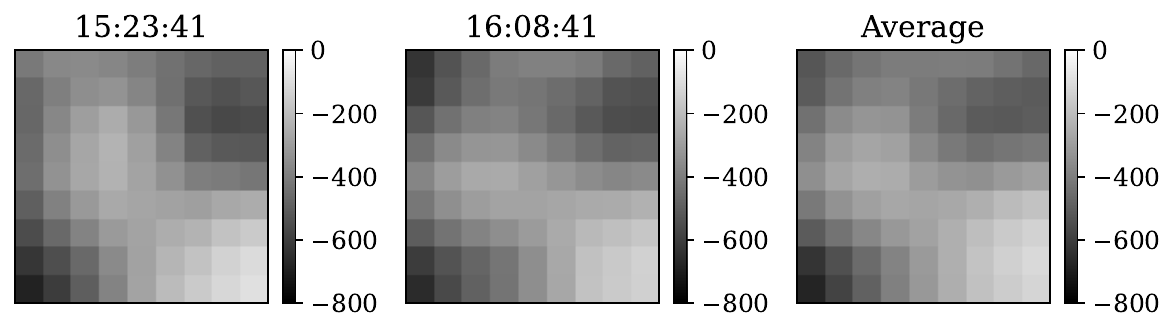}
    \caption{The left and middle panels show the plume footpoint in HMI LOS at the start and end of this coordinated observation (white dotted box in Figure \ref{fig:contour}). The right panel gives the average pixel value over the course of the observation. The color bar is not centered on zero in order to highlight the differences in this region, since there are no positive field pixels.
    }
    \label{fig:start_end_B_field}
\end{figure}

We observe no signatures of mixed-polarity flux at the footpoint, but we are constrained by HMI's limited sensitivity and resolution (Figure \ref{fig:start_end_B_field}). Yet mixed-polarity flux in sunspot penumbra has been observed in high spatial and spectral resolution observations \citep{Joshi_2011,Scharmer_2012,Tiwari_2013,Tiwari_2015,Esteban_Pozuelo_2015}, so we cannot rule out the possibility that there is mixed-polarity magnetic field present in this region of the penumbra.

\vfill\null
\section{Animation}
\label{sec:animation}
\begin{figure}[H]
    \centering
    \includegraphics[width=\linewidth]{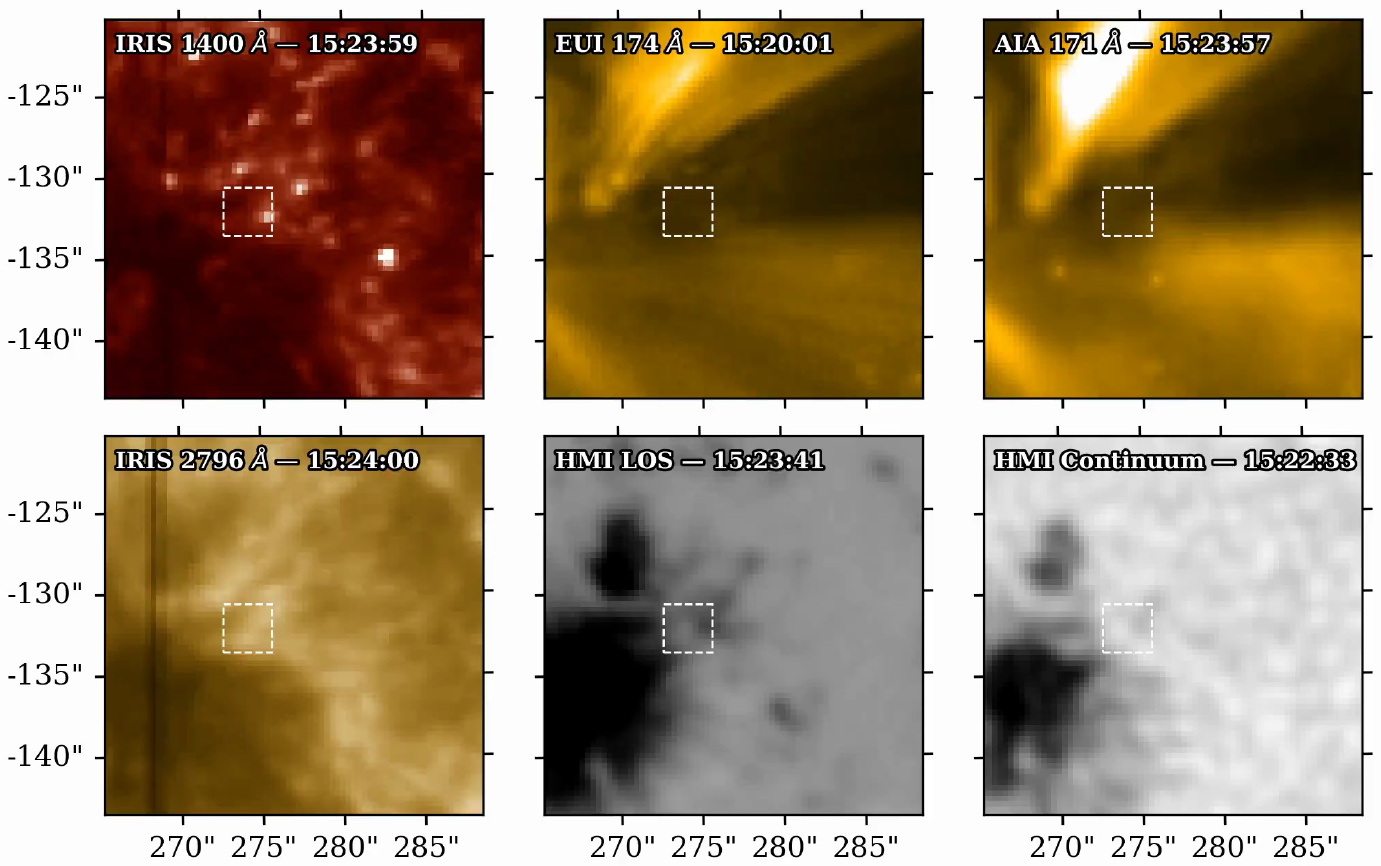}
    \caption{Zoomed-in region from Figure \ref{fig:context} where the plume of interest develops. Top row left to right: IRIS 1400 \AA, EUI 174 \AA, AIA 171 \AA. Bottom row left to right: IRIS 2796 \AA, HMI LOS, HMI continuum. The dashed white box indicates the approximate position of the plume footpoint (same position as in Figure \ref{fig:contour}). Animation spans the entire EUI observation (15:24:01—16:09:20 UT). The color bars used are the same as those used in Figure \ref{fig:contour}. Channels not shown in Figure \ref{fig:contour}, AIA 171 \AA{} and IRIS 1796 \AA, use the following ranges: $0-7$ (same as EUI 174 \AA) and $0-4$, respectively. (An animation of this figure is available in the online article.)
    }
    \label{fig:animation}
\end{figure}

Figure \ref{fig:animation} is a still from Figure \ref{fig:context}'s corresponding animation.

\bibliographystyle{aasjournal}
\bibliography{references}

\end{document}